\documentclass[12pt]{article}

\usepackage{amsmath,amssymb,amsthm,graphicx,graphics}
\usepackage{fullpage}

\begin{document}

\title{Thoughtful comments on \ `Bessel beams and signal propagation' \ 
}
\author{E. Capelas de Oliveira$^{*}$, W. A. Rodrigues, Jr.$^{*,\star }$, 
D. S. Thober$%
^{\star }$ \\ and  \\ A. L. Xavier, Jr.$^{\star }$ \\
$^{*}$ Institute of Mathematics, Statistics and Scientific 
Computation,\\
{\em IMECC-UNICAMP}\\
CP 6065, 13083-970, Campinas, SP, Brazil\\
$^{\star }$Wernher Von Braun - Advanced Research Center, Unisal\\
Av. A. Garret, 267, 13087-290, Campinas, SP, Brazil}
\date{\today }
\maketitle

\begin{abstract}
In this paper we present thoughtful comments on the paper \ `Bessel 
beams
and signal propagation' showing that the main claims of that paper are
wrong. Moreover, we take the opportunity to show the non trivial and 
indeed
surprising result that a scalar pulse (i.e., a wave train of compact 
support
in the time domain) that is solution of the homogeneous wave equation (
vector ($\vec{E},\vec{B}$) pulse that is solution of Maxwell equations) 
is
such that its {\em wave front} in some cases does travel with speed \ 
{\em %
greater} than $c$, the speed of light . In order for a pulse to posses a
front that travels with speed $c$, an additional condition must be
satisfied, namely the pulse must have finite energy. When this condition 
is
fulfilled the pulse still can show peaks propagating with superluminal 
(or
subluminal) velocities, but now its wave front travels at speed $c$. 
These
results are important because they explain several experimental results
obtained in recent experiments, where superluminal velocities have been
observed, without implying in any breakdown of the Principle of 
Relativity.
\end{abstract}

In this paper we present some thoughtful comments ({\bf C}$_{1}-{\bf 
C}_{4})$
concerning statements presented in the paper \ \ `Bessel beams and 
signal
propagation' [1] and also some non trivial results concerning 
superluminal
propagation of peaks in particular electromagnetic pulses in 
nondispersive
media.

In [1] the author recalls that the experimental results presented in [2]
showed that Bessel \ beams generated at microwave frequencies have a 
{\em %
group velocity }greater than the velocity of light $c$ (in what follows 
we
use units such that $c=1$)\footnote{%
In [3] we scrutinized the experimental results of [2]. We presented 
there a
simple model showing that all particulars of the data (including the 
slowing
of the superluminal velocity of the peak along the propagation 
direction)
can be qualitatively and quantitatively understood as a scissor's like
effect. Moreover in [3] we called the readers attention that in [4] 
peaks of
{\em finite aperture approximations (FAA)} to particular acoustical 
Bessel
pulses called $X$-waves (first discoverd by Lu and Greenleaf ([5,6]) 
have
been see to travel at supersonic speed i.e., with velocity greater than 
$%
c_{s\text{ }}$, the sound speed parameter appearing on the homogenous 
wave
equation ({\em HWE}). In [4] and [7] it is also predicted the possibilty 
of
launching \ {\em FAA} to superluminal electromagnetic $X$-waves, a fact 
that
has been confirmed experimentally in the microwave region in [2] \ and 
in
the optical region in [8]. A review concerning the different facets of
`superluminal' wave motion under different physical conditions can be 
found
in [9].}. His intention was then to show that the signal velocity, 
defined
according to Brillouin and Sommerfeld ({\em B\&S}) was also 
superluminal. We
explicitly shows that the particular example used by the author of [1],
given by the Bessel beam of his eq.(3) does not endorse his claim. 
Contrary
to the author's conclusion this beam has no fronts in both space and 
time
domains, hence cannot satisfy {\em B\&S} defintion of a signal. 
Moreover,
the beam given by eq.(3) of [1] travels rigidly with a superluminal 
speed.
We prove then that there are two classes of general Bessel pulses 
satisfying
{\em B\&S} definition of signal. A solution of the {\em HWE} 
corresponding
to class I is such that the group speed is always less than $c$ \ 
whereas
its front moves with speed $c$.\footnote{%
Of course, this is a kind of generalized reshaping phenomena which 
cannot
endures for ever. It lasts until the peak of the wave catches the 
front.} A
solution of the {\em HWE} of the class II travels rigidly at 
superluminal \
speed if care is not taken of the energy content of the pulse. We 
present
also some necessary comments concerning solutions of Maxwell equations
associated with Bessel beams of classes I and II.

We start by recalling the general solution of the {\em HWE} $\
\square \Phi =0$ in Minkowski spacetime ($M,\eta ,D$) [10-12]. In
a given Lorentz reference frame [10-12] $I=\partial /\partial t\in
\sec TM$, we choose cylindrical coordinates ($\rho ,\varphi ,z$)
naturally adapted to the $I$ reference frame, where $\rho
=(x^{2}+y^{2})^{\frac{1}{2}}$ and $x=\rho \cos \varphi $ and
$y=\rho \sin \varphi $, with ($x,y,z$) being the usual cartesian
coordinates naturally adapted to $I$. Writting
\begin{equation}
\Phi (t,\rho ,\varphi ,z)=f_{1}(\rho )f_{2}(\varphi )f_{3}(t,z),
\end{equation}
and substituting eq.(1) in the {\em HWE} we get the following equations
(where $\nu $ and $\Omega $ are {\em separation parameters}),
\begin{equation}
\left\{
\begin{array}{c}
\left[ \rho ^{2}\frac{d^{2}}{d\rho ^{2}}+\rho \frac{d}{d\rho }+(\rho
^{2}\Omega ^{2}-\nu ^{2})\right] f_{1}=0, \\
\left( \frac{d^{2}}{d\varphi ^{2}}+\nu ^{2}\right) f_{2}=0, \\
\left( \frac{\partial ^{2}}{\partial t^{2}}-\frac{\partial 
^{2}}{\partial
z^{2}}+\Omega ^{2}\right) f_{3}=0.
\end{array}
\right. .  
\end{equation}

The first of eqs.(2) is Bessel's equation, the second one implies that 
$\nu $
must be an integer and the third is a Klein-Gordon equation in two
dimensional Minkowski spacetime.\footnote{%
In 4-dimensional spacetime the Klein-Gordon equation possess families of
luminal and superluminal solutions, besides subluminal solutions. See 
[4]
and references therein.} In what follows ( without loss of generality 
for
the objectives of the present paper) we choose $\nu =0$ (and also 
$\Omega >0$%
). Then, \ we obtain as a solution of eqs.(2) a wave propagating in the 
$z$%
-direction, i.e.,
\begin{equation}
\Phi _{J_{0}}(t,\rho ,z)=J_{0}(\rho \Omega )\exp [-i(\omega 
t-\bar{k}z)],
\end{equation}
where the following dispersion relation must necessarily be satisfied,
\begin{equation}
\omega ^{2}-\bar{k}^{2}=\Omega ^{2}.  
\end{equation}

The dispersion relation given by eq.(4) may look strange at first sight, 
but
there are evidences that it can be realized in nature (see below) in 
some
special circunstances.

{\bf C}$_{1}$. It is quite clear that the wave described by
eq.(3),
called in [1] a Bessel beam\footnote{%
Note that in [1] the author writes $\Omega =\omega \sin \theta $ and 
$\bar{k}%
=\omega \cos \theta $.}, has phase velocity $v_{ph}=\omega 
/\bar{k}>1$.
However, we point out that the statement done in [1] is {\em false}, 
namely:
`As known, in the absence of dispersion the group velocity $v_{gr}$ of a
Bessel pulse is equal to the phase one [4,5]\footnote{%
The references [4,5] in [1] are the references [8,13] in the present 
paper.}
since all the components at different frequencies propagate with the 
same
velocity'. To prove its falsity recall that there exists a Lorentz 
reference
frame [10-12]
\begin{equation}
I^{\prime }=(1-v_{gr}^{2})^{\frac{1}{2}}\left( \partial /\partial
t+v_{gr}\partial /\partial z\right) \in \sec TM,  
\end{equation}
which is moving with velocity $v_{gr}=d\omega /d\bar{k}<1$ in relation 
to
the frame $I$ in the $z$-direction. In the coordinates naturally adapted 
to
the frame $I^{\prime }$ the frequency of the wave is $\omega ^{\prime 
}=$ $%
\Omega $, which means that in the frame $I^{\prime }$ the Bessel beam is
stationary. This proves our statement that for Bessel beam the group
velocity is always less than the velocity of light $c.$

{\bf C}$_{2}$. Now, we show how \ to build two different classes (I and 
II)
of solutions of the {\em HWE} by appropriate linear superpositions of 
waves
of the form given by our eq.(3).

{\bf Class I}. Suppose, following {\em B\&S} [13,14 ] that a signal is
defined as a pulse with a {\em finite} time duration at the origin 
$z=0$
where a physical device generated it. We model our problem as a {\em %
Sommerfeld problem} [15] for the {\em HWE }(with cylindrical
symmetry), i.e., we want to find the solution of the {\em HWE}
with the following conditions (called in what follows Sommerfeld
conditions),
\begin{eqnarray}
\Phi (t,\rho,0) &=&AJ_{0}(\rho \Omega )[\Theta (t)-\Theta
(t-T)]\sin \omega _{0}t\text{ }  \nonumber \\
&=&AJ_{0}(\rho \Omega )\frac{1}{2\pi }\Re  \int_{\Gamma }d\omega {\rm 
e}%
^{-i\omega t}\frac{\left\{ {\rm e}^{i\omega T}-1\right\} }{\omega 
-\omega
_{0}},  \nonumber \\
\left. \frac{\partial \Phi (t,\rho ,z)}{\partial z}\right| _{z=0}
&=&AJ_{0}(\rho \Omega )\frac{1}{2\pi }\Re  \int_{\Gamma }d\omega 
\text{
}\bar{k}(\omega )\text{ }{\rm e}^{-i\omega t}\frac{\left\{ {\rm 
e}^{i\omega
T}-1\right\} }{\omega -\omega _{0}}.  
\end{eqnarray}

In eq.(6) $\Theta (t)$ is the Heaviside function, $A$ and $\omega
_{0}=\Omega $ are constants, $\Re$ means real part and
$\bar{k}(\omega )$ is given below and for simplicity we take
$T=N\tau _{0}=2\pi N/\omega _{0}$, with $N$ an integer. Now, to
solve our problem it is enough to get a solution of the third of
eqs.(2). We have,
\begin{equation}
f_{3}(t,z)=\frac{1}{2\pi }\Re \int_{\Gamma }\frac{d\omega }{\omega
-\omega _{0}}\left\{ {\rm e}^{-i\omega (t-T-v_{gr}z)}-{\rm e}^{-i\omega
(t-v_{gr}z)}\right\}   
\end{equation}
where $v_{gr}=\bar{k}(\omega )/\omega $ and $\Gamma $ is an 
appropriate path
in the complex $\omega $-plane. We note $\lim_{\omega \rightarrow \infty
}v_{gr}=1.$ Putting eq.(7) into the third of eqs.(2) we see that the
dispersion relation given by eq.(4) must be satisfied. To continue we 
write,
\begin{equation}
\bar{k}(\omega )=\sqrt{(\omega +\Omega )(\omega -\Omega )}.  
\end{equation}


\begin{figure}[htb]
\centering
\includegraphics{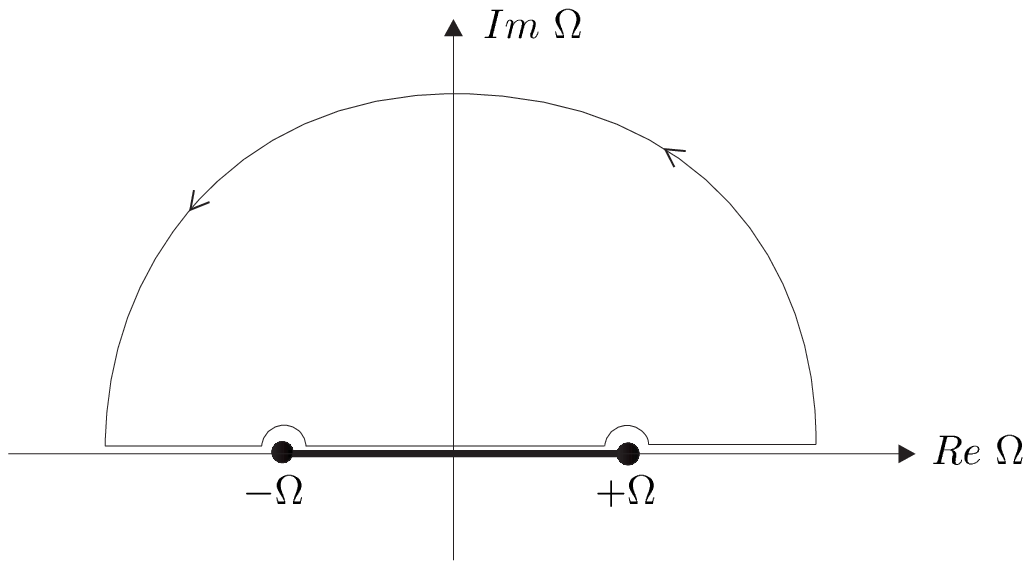}
\caption{Contour for integration of eq.(7) for $t-z<0$}
\end{figure}

There are two branch points at $\omega =\pm \Omega $. The 
corresponding
branch cuts can be taken as the segment ($-\Omega ,\Omega $) in the real 
$%
\omega $-axis. Following $\Gamma $ from positive values of $\Re \omega $
above and close to the real axis, the root in eq.(8) acquires a phase 
factor
$e^{i\pi }=-1$ when passing from $\Re \omega >\Omega $ to $\Re %
\omega <-\Omega $. Then, on the real $\omega $-axis we have,
\begin{equation}
\bar{k}(\omega )=\left\{
\begin{array}{c}
|\sqrt{\omega ^{2}-\Omega ^{2}}|,\hspace{0.1in}\omega >\Omega \\
-|\sqrt{\omega ^{2}-\Omega ^{2}}|,\hspace{0.1in}\omega <-\Omega
\end{array}
\right.  
\end{equation}
a result that is necessary in order to calculate the value of $f_{3}$ 
for $%
(t-v_{gr}z)>0$. We are not going to investigate this case here, since we 
are
interested in the behavior of $f_{3}$ for the case where $(t-z<0)$. In 
this
case, we must close the contour $\Gamma $ in the upper half plane. Since
there are no poles inside the contour we get that
\begin{equation}
f_{3}(t,z)=0\hspace{0.2in}\text{ for\hspace{0.2in} }t-z<0.  
\end{equation}

Now, it is easy to verify the intensity of the wave which is solution of 
the
{\em HWE} and satisfies the Sommerfeld conditions given by eq.(6) has a
maximum for $\omega =\omega _{0}$, i.e., the waves with frequency near 
$%
\omega _{0}$ have always a much greater amplitude than all others. Under
these conditions let us write,
\begin{equation}
\omega t-\bar{k}z=(\omega 
_{0}t-\bar{k}_{0}z)+(t-\frac{z}{v_{gr0}})(\omega
-\omega _{0}),  
\end{equation}
where $v_{gr0}=(d\omega /d\bar{k})|_{\omega =\omega _{0}}<1$ and $%
v_{ph0}=\omega _{0}/\bar{k}_{0}>1$. We can write an approximation for 
the
function $f_{3}(t,z)$ denoted by $\tilde{f}_{3}(t,z)$ as,
\begin{equation}
\tilde{f}_{3}(t,z)=\frac{1}{2\pi }\Re  \left\{ {\rm e}^{-i\omega
_{0}(t-z/v_{ph0})}\int\limits_{\omega _{0}-\triangle \omega }^{\omega
_{0}+\triangle \omega }\frac{d\omega }{\omega -\omega _{0}}\left\{ {\rm 
e}%
^{-i\omega (t-T-z/v_{gr0})}-{\rm e}^{-i\omega (t-z/v_{gr0})}\right\}
\right\} .  
\end{equation}

We see that $\tilde{f}_{3}(t,0)$ is equal to $f_{3}(t,0)$ if we suppress 
in
the expression for this function the frequencies very different from 
$\omega
_{0}$. Now, $\tilde{f}_{3}(t,0)$ has support on the whole temporal axis,
i.e., in the interval $-\infty <t<\infty $, but it is taken by some 
authors
(like, e.g., [16]) as representing a wave that begin gradually at 
$t=0$ and
ends gradually at $t=T$. Of course, no wave of the kind of 
$\tilde{f}_{3}$
can be build by any physical device. The importance of the function 
$\tilde{f%
}_{3}(t,z)$ is that, as emphasized by {\em B\&S} [13,14] it shows that 
we
can associate a {\em group velocity} to pulse peaks in general (and of
Bessel beams in particular) satisfying the Sommerfeld conditons (eq.(6)) 
and
that the group velocity in this case is {\em less} than the velocity of
light. This means that after a while the {\em back} end of the wave that 
is
travelling at speed $c(=1)$ will catch the peak. The wave reshapes 
even when
propagating in vacuum.

A general subluminal $J_{0}$-Bessel beam can be written as,
\begin{equation}
\Phi _{{B}}(t,\rho ,z)=J_{0}(\rho \omega ){\cal F}^{-1}[T(\omega 
)]{\rm e%
}^{i\bar{k}z}  
\end{equation}
where $T(\omega )$ is an appropriate transfer function and ${\cal 
F}^{-1}$
is the inverse Fourier transform. \ Now, the {\em peaks} of {\em FAA} to 
\
acoustical pulses of the form given by eq.(13) (i.e., the waves at 
$z=0$ are
not zero only in the time interval $0<t<T$) have been seen travelling at
{\em subluminal} speed\footnote{%
Of course, in this case the speed paramenter appearing in the {\em HWE} 
must
be $c_{s}$, the sound speed in the medium, and the word subluminal speed
used must be understood as a speed less than $c_{s}$.} in an experiment
described in [4], thus endorsing the above analysis.

{\bf Class II}. We now return to the dispersion relation given by eq.(4) 
and
write,
\begin{equation}
\bar{k}=k\cos \theta ,\hspace{0.2in}\Omega =k\sin \theta ,  
\end{equation}
where $\theta $ is a constant called {\em axicon} angle [5,6,17]. It 
results
that
\begin{equation}
\omega =\pm k.  
\end{equation}

We immediately verify that
\begin{equation}
J_{0}(\omega \rho \sin \theta ){\rm e}^{-i(\omega t-kz\cos \theta )},
\end{equation}
is a solution of the {\em HWE} whose beam width is proportional to 
$1/\omega
\sin \theta $, thus being frequency dependent. The dependency of the 
beam
width on frequency will cause the beam to have a pulse response that is
independent of position. Indeed, suppose that the source is driven by a
frequency distribution $B(\omega )$, i.e., we have a pulse
\begin{equation}
\Phi _{X}(t,\rho ,z)=\int\limits_{-\infty }^{\infty }d\omega B(\omega
)J_{0}(\omega \rho \sin \theta ){\rm e}^{-i(\omega t-kz\cos \theta 
)},\text{
}\omega =k.  
\end{equation}

If $J_{0\text{ \ }}$were not dependent on frequency the integral in 
eq.(17)
would be simply the inverse Fourier transform of the source spectrum and 
we
return to class I solutions. However, here $J_{0}$ is dependent on 
frequency
and also on position and consequently modifies the pulse spectrum in 
such a
way to make the time response of the pulse dependent on radial position. 
We
put an index $X$ in the wave given by eq.(17) because pulses of this 
kind
have been named $X$-waves by Lu and Greenleaf since 1992 [5,6]. Even 
more,
taking $B(\omega )=Ae^{-a_{0}|\omega |}$ ($A$ and $a_{0}>0$ being
constants), we can easily verify (c.r., pages 707 and 763 of [18]) that 
we
can write for $\sin \theta >0$,
\begin{subequations}
\begin{equation}
\Phi_{X}(t,\rho ,z) = A\int_{-\infty }^{\infty }d\omega {\rm e}%
^{-a_{0}|\omega |}J_{0}(\omega \rho \sin \theta ){\rm e}^{-i\omega 
(t-z\cos
\theta )}  
\end{equation}
\begin{equation*}
= A\int_{0}^{\infty }d\omega {\rm e}^{-a_{0}\omega }J_{0}(\omega
\rho \sin \theta )\cos (\omega \mu )
\end{equation*}
\begin{equation}
= \frac{A}{\left[ \rho ^{2}\sin ^{2}\theta +\left[ a_{0}+i\mu \right]
^{2}\right] ^{\frac{1}{2}}}+\frac{A}{\left[ \rho ^{2}\sin ^{2}\theta 
+\left[
a_{0}-i\mu \right] ^{2}\right] ^{\frac{1}{2}}}  
\end{equation}
\begin{equation*}
= \frac{A\sqrt{2}\left\{ \left[ \left[ \rho ^{2}\sin ^{2}\theta
+a_{0}^{2}-\mu ^{2}\right] ^{2}+4a_{0}^{2}\mu ^{2}\right] ^{\frac{1}{2}%
}+\rho ^{2}\sin ^{2}\theta +a_{0}^{2}-\mu ^{2}\right\} ^{\frac{1}{2}}}{%
\left\{ \left[ \rho ^{2}\sin ^{2}\theta +a_{0}^{2}-\mu ^{2}\right]
^{2}+4a_{0}^{2}\mu ^{2}\right\} ^{\frac{1}{2}},}  
\end{equation*}
\begin{equation}
\end{equation}
\end{subequations}
where $\mu =($ $t-z\cos \theta )$.

Eq.(18c) shows that this wave is a real solution of the {\em HWE}. We 
recall
that if in eq.(18a) we use as integration interval $0<\omega <\infty $, 
we
get only the first term in eq.(18b). In this case we have a complex wave
that has been called the {\em broad band} \ $X$-wave in [4-6]. These 
waves
and the more general ones given by eq.(18b) propagate without distortion
with superluminal velocity given by $1/\cos \theta $, but of course they
cannot be produced in the physical world because (like the plane wave
solutions of the {\em HWE}) they have {\em infinity energy}, as it is 
easy
to verify. Waves that are solutions of the linear relativistic wave
equations and that propagate in a distortion free mode, have been called
{\em UPWs} (undistorted progressive waves) in [4].

Now, we show that a X-pulse even if it has compact support in the time
domain (thus being of the form of a {\em B\&S} signal) is such that its
front propagates with superluminal speed. To prove our statement we look 
for
a solution of the {\em HWE} satisfying the following Sommerfeld 
conditions%
\footnote{$B(\omega )$ is taken in this example as a function such that 
$%
\int\limits_{-\infty }^{\infty }d\omega B(\omega )J_{0}(\omega \rho \sin
\theta ){\rm e}^{-i\omega t}$ has support in the interval $-\infty 
<t<\infty
$.},

\begin{equation}
\begin{array}{rcl}
\Phi _{X}(t,\rho ,0) &=&\left[\Theta (t+T)-\Theta (t-T) \right]
\int\limits_{-\infty }^{\infty }d\omega B(\omega )J_{0}(\omega \rho \sin
\theta ){\rm e}^{-i\omega t}, \\
\left. \displaystyle\frac{\partial \Phi_X (t,\rho ,z)}{\partial z}\right| _{z=0} 
&=&i\left[
\Theta (t+T)-\Theta (t-T)\right] \cos \theta\int\limits_{-\infty }^{\infty }d\omega
B(\omega )J_{0}(\omega \rho \sin \theta ) k(\omega) {\rm e}^{-i\omega t}, 
\end{array}
\end{equation}
and $k(\omega )=\omega $. Proceeding in the same way as in the 
Sommerfeld
problem of class I solution presented above we obtain as a solution of 
the
{\em HWE} (for $z>0$),
\begin{eqnarray}
\Phi _{X}(t,\rho ,z) &=&\frac{1}{2\pi }\int\limits_{-\infty }^{\infty 
}d\bar{%
\omega}B(\bar{\omega})J_{0}(\bar{\omega}\rho \sin \theta
)\int\limits_{-\infty }^{\infty }d\omega {\rm e}^{-i\omega (t-z\cos 
\theta
)}\left[ \frac{{\rm e}^{i(\omega -\bar{\omega})T} - {\rm e}^{-i(\omega -\bar{\omega})T}}{i(\omega 
-\bar{\omega})}%
\right]   \nonumber \\
&=&\left\{
\begin{array}{l}
\int\limits_{-\infty }^{\infty }d\omega B(\omega )J_{0}(\omega \rho \sin
\theta ){\rm e}^{-i\omega (t-z\cos \theta 
)}\hspace{0.1in}\text{for\hspace{%
0.1in}}\left| t-z\cos \theta \right| <T \\
\hspace{1in}0\hspace{1.1in}\hspace{0.1in}\text{for\hspace{0.1in}}\left|
t-z\cos \theta \right| >T
\end{array}
\right. .  \nonumber \\
&&  
\end{eqnarray}


\begin{figure}[htb]
\centering
\includegraphics{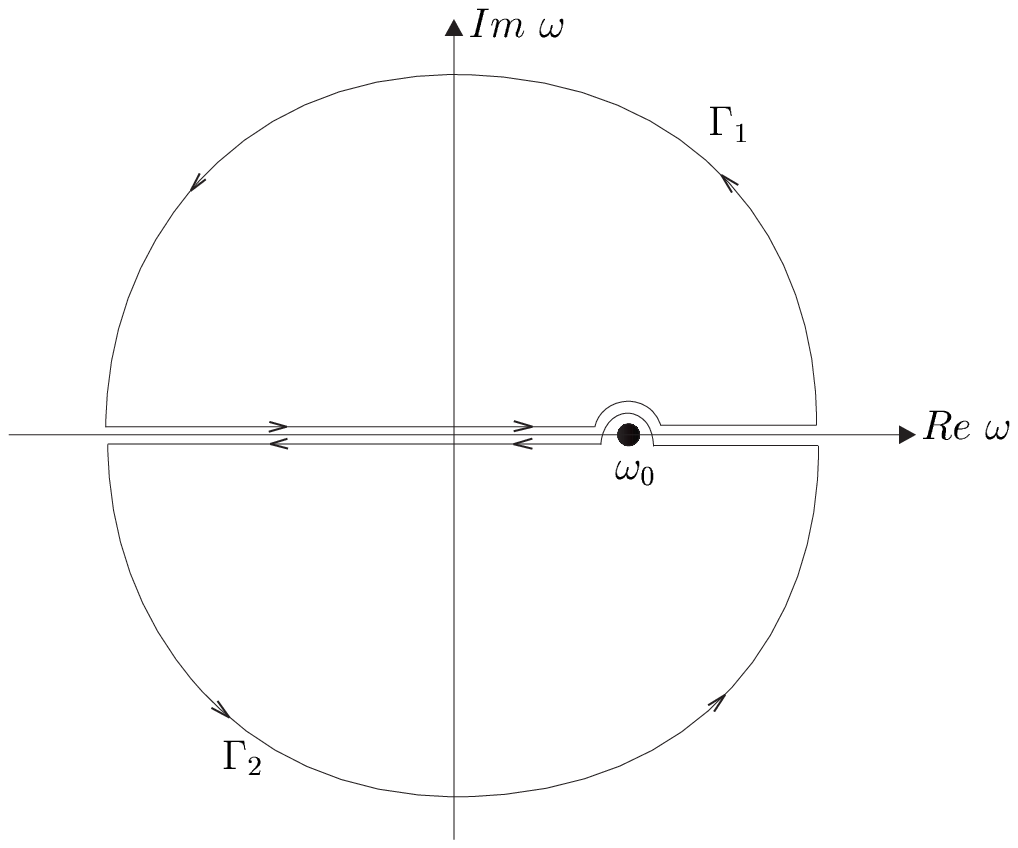}
\caption{Contours for integration of
eq.(20). $\Gamma _{1}$ for $\left| t-z\cos \theta \right| >T$ and
$\Gamma_{2}$ for $\left| t-z\cos \theta \right| <T$.}
\end{figure}

We see that for $\left| t-z\cos \theta \right| <T$ the integral in 
eq.(20) is%
{\em \ not} zero. Since the axicon angle $\theta >0$, then $1>\cos 
\theta < 0$
and it follows that the pulse is not zero for $z>t$ and $t>T$, what 
means
that the wave front of our pulse propagates with {\em superluminal} 
speed!
Of course, the pulse is zero for $z<(t-T)/\cos \theta $ or $z>(t+T)/\cos
\theta $. We observe that the above result is true even a single Bessel
pulse, i.e., when $B(\omega )=\delta (\omega -\omega _{0})$, a result 
that
we mentioned in [3].

How to compare this finding with the famous {\em B\&S} result
[13,14] stating that a wave pulse which propagates in a dispersive
medium with loss has a front propagating at maximum speed $c$?
Some things are to be recalled in order to get a meaningful
answer. The first is that {\em B\&S} example refers to a
propagation of a `plane' wave truncated in time (which, of course,
has {\em infinite} energy) satisfying the Sommerfeld conditions
(analogous to eq.(6)) and propagating in a dispersive medium with
loss. A careful analysis [19] shows that the same problem in a
dispersive medium with gain reveals that in this case we can find
two kinds of solutions ( both of of infinite energy). In one of
these kinds, by appropriately choosing the integration path in the
complex $\omega $-plane we obtain as result that the front of the
wave may travel with superluminal speed. This situation is
somewhat analogous to what happen with some possible mathematical
solutions of the tachyonic Klein-Gordon equation in two
dimensional Minkowski spacetime [20,21]. This equation is
important because it can be associated with the so called
telegraphist equation.

The reason for our finding that the $X$-pulse propagating in a 
nondispersive
medium, although of compact support in the time domain, is such that its
front travel at superluminal speed is the following; the solution given 
by
eq.(20) is not of {\em compact support} in the space domain and as such 
has
infinite energy as can be easily verified. Only for a pulse of finite 
energy
we can warranty that its front always travel with a speed that cannot be
greater than maximum speed. Indeed, suppose we produce on the plane 
$z=0$ a
pulse like the one given by eq.(20), except that it has a finite lateral
circular width of radius $a$, i.e.,it is taken as zero for $\rho >a$. 
Such a
pulse is called a {\em FAA} to the pulse given by eq.(20) and as can be
easily verified has {\em finite energy}. \ If such a pulse does not 
spread
with infinite velocity during its build up, then after it is ready, 
i.e., at
$t=T$ it occupies a region of compact support in space given by 
$\left| \vec{%
x}\right| <R$, where $R$ is the maximum linear dimension involved. Such 
a
field configuration can then be taken as part of the initial conditions 
for
a {\em strictly} hyperbolic Cauchy problem at $t=T$. For such a 
problem it
is well known the mathematical theorem that stablishes that [22,23] \ 
the
time evolution of the pulse must be such that it is {\em null} for 
$\left|
\vec{x}\right| >R+c(t-T)$. In conclusion, it is not sufficient for a 
wave to
be of compact support in the time domain (i.e., to be a pulse) to assure
that the wave front of the pulse moves in a nondispersive medium at 
maximum
speed $c$. In order for the wave front to move with velocity $c$ it is
necessary that the pulse possess {\em finite energy}, and in order for 
this
condition to be satisfied the pulse must have compact support in the 
space
domain after its build up. We recall here that in [4] the peaks of {\em 
FAA}
to acoustical pulses given by eq.(18) (with appropriated $B(\omega )$) 
have
been seen traveling with velocities $c_{s}/\cos \theta $, thus 
confirming
the theory developed above.

${\bf C}_{3}$. We now examine the claim of [1] that a wave given by our
eq.(17), with $B(\omega )=1$, i.e.,
\begin{equation}
U(t,\rho ,z)=\int\limits_{-\infty }^{\infty }d\omega J_{0}(\omega \rho 
\sin
\theta ){\rm e}^{-i(\omega t-kz\cos \theta )},\text{ }\omega =k.
\end{equation}
is a pulse with support only in the $z$-axis at points $z=t/\cos 
\theta $
and with value at that points $\delta (0)$. The calculations presented 
in
[1] are wrong. Before we prove our statement let us recall that [1] 
quotes
Brillouin: \ `a signal can be defined as a pulse of finite temporal
extension, that is, of infinite extension in the frequecy 
domain'.\footnote{%
This definition is due to Sommerfeld. See [13,14].} The wave given by
eq.(21) has an infinite extension in the frequency domain but it is not 
a
pulse of finite time domain (for a fixed $z$). Indeed, as theorem 11 on 
page
22 in Sneddon's book [24] stablishes: a function which is bounded in the
time domain has an infinite extension in the frequency domain, but it is 
not
true that a function with an infinite frequency spectrum is necessarily
bounded in the time domain. A trivial example of the last statement is 
the
case of a Gaussian pulse, whose Fourier transform is itself a Gaussian. 
In
the particular case of the wave given by eq.(21) it is immediate to 
realize
that the integral is nothing more than the{\em \ Fourier transform} of a 
$%
J_{0}$ function, and the value of the integral is given in many books, 
in
particular on page 523 of Sneddon's book [24]. We have,
\begin{subequations}
\begin{equation}
\int\limits_{-\infty }^{\infty }d\omega J_{0}(\omega \rho \sin \theta )%
{\rm e}^{-i(\omega t-kz\cos \theta )}  
\end{equation}
\begin{equation}
= \left\{
\begin{array}{c}
\frac{2}{\sqrt{\rho ^{2}\sin ^{2}\theta -(t-z\cos \theta )^{2}}}\text{ 
for }%
|t-z\cos \theta |<\rho \sin \theta \text{ } \\
0\text{ for }|t-z\cos \theta |>\rho \sin \theta
\end{array}
\right.  
\end{equation}
\end{subequations}

Eq.(22b) shows that $U(t,\rho ,z)$ has support in the entire time axis
provided that $|t-z\cos \theta |<\rho \sin \theta $. When $\rho =0$, 
since $%
U $ is real (as can be seen directly from eq.(22a) we must have
that $|t-z\cos \theta |=0$ and the function $U$ is singular. We see 
that the
result expressed by eq.(22b) is compatible with the one given by 
eq.(18b) if
we take the limit for $a_{0}\rightarrow 0$.

{\bf C}$_{4}$. Finally, we investigate the claim (done in [1] \ and
attributed to [8]) that the\ wave function given by eq.(3) represents an
electric field. This claim is a nonsequitur. Indeed, \ the scalar 
solutions
of the {\em HWE} can be used to generated solutions of the Maxwell 
system
using the {\em Hertz} potential method (see, e.g.[25,26]). In 
particular,
superluminal solutions of the {\em HWE} can be used to produce 
superluminal
solutions of Maxwell equations [4,7,9]. If we choose a {\em magnetic} 
Hertz
potential $\vec{\Pi}_{m}=\Phi _{J_{0}}\hat{z}$ it is a simple exercise 
to
show that the transverse electric and magnetic fields do not show any
dependence on $J_{0}$. Only the $B_{z}$ component of the electromagnetic
field configuration has a $J_{0}$ dependence.
Explicitly we have from the well
known formulas $\vec{E}=-\partial /\partial t(\nabla \times 
\vec{\Pi}_{m})$
and $\vec{B}=\nabla \times \nabla \times \vec{\Pi}_{m}$ that,
\begin{eqnarray}
E_{\rho } &=&0,\text{ }E_{\varphi }= i\omega \Omega 
J_{1}(\Omega
\rho ){\mbox{e}}^{-i(\omega t-\bar{k}z)},\text{ }%
E_{z}=0,  \nonumber \\
 B_{\rho} & = & -k \Omega J_1 (\Omega \rho) {\mbox{e}}^{-i(\omega t -
 \bar{k}z)} \nonumber ,\\
B_{z} &=& \Omega^2 \left[ \frac{J_{1}(\Omega \rho 
)}{\Omega \rho} + \frac{J_{0}(\Omega \rho )}{2} - \frac{J_{2}(\Omega \rho
)}{2} \right] {\mbox{e}}^{-i(\omega t - {\bar{k}z})} , \nonumber \\
\omega ^{2}-\bar{k}^{2} &=&0.  
\end{eqnarray}

With an electric Hertz potential we obtain a solution where only the 
$E_{z}$
component \ has a $J_{0}$ dependence. As such, we conclude that the
electromagnetic beams observed in [2] and also in [8,17] are not $J_{0}$
beams. A careful analysis of the solutions of Maxwell equations in
cylindrical symmetry shows that there are not $J_{0}$ solutions 
representing
transverse electric fields. The existence of only one peak observerd in 
the
experiments done in [2] must be due to the $J_{1}/\rho $ term in 
$E_{\varphi
}$. A more detailed analysis will be reported elsewhere.

Our conclusions are as follows: (i) our results show that the main 
claims of
[1] are wrong and/or misleading and leads to equivocated conclusions
concerning recent experimental results showing superluminal motion of 
peaks
of particular electromagnetic field configurations in {\em 
nondispersive}
media; (ii) we also prove a non {\em trivial} result, namely that the
condition that a wave is of {\em finite} time duration is not a {\em %
sufficient} condition for its front to propagate with the speed $c$. It 
is
necessary in order for the front to travel with speed $c$ that the pulse
possess {\em finite} energy, \ and thus as explained above it must 
(after
being prepared by the launching device) have support only in a compact 
space
region when ready;\footnote{%
We mention here that any electromagnetic pulse fulfilling this condition
spreads, a result that may be called the non focusing theorem [27].} 
(iii)
only {\em FAA} to superluminal solutions of the {\em HWE} (acoustical 
case)
and to superluminal solutions of Maxwell equations can be produced in
nature, because only waves of this kind have finite energy. These {\em 
FAA }%
exhibit peaks propagating with superluminal speeds even in the vacuum, 
but
since their fronts propagate with speed $c$ this kind of phenomenom does 
not
implies in any danger for the Theory of Relativity.\bigskip 

Acknowledgments: W.A.R., D.S.T. and A.L.X.Jr. are grateful to Motorola
Industrial Ltda. for a research grant. A. L. X. Jr. would like also to thank
FAPESP (Funda\c{c}\~ao de Amparo \`a Pesquisa do Estado de S\~ao Paulo)
 for financial support under contract 00/03168-0. The authors are also 
grateful to Dr.
J. E. Maiorino and Professor J. Vaz Jr. for useful discussions.\bigskip

{\bf References\bigskip }

[1] D. Mugnai, {\em Bessel beams and signal propagation}, in publication 
in
{\em Phys. Lett A} (2001).

[2] D. Mugnai, A. Ranfagni and R. Ruggeri, {\em Phys. Rev. Lett.} {\bf 
80}
(2000) 4830.

[3] W. A. Rodrigues Jr, D. S. Thober and A. L. Xavier, {\em Causal
explanation of observed superluminal behavior of microwave propagation 
in free space}, http://arXiv.org/abs/physics/0012032, (2001) in publication in Phys. Lett. A (PLA10711).

[4] W. A. Rodrigues, Jr. and J. Y. Lu, {\em Found. Phys.} {\bf 27} 
(1997)
435.

[5] J.Y. Lu and J. F. Greenleaf, {\em IEEE Trans. Ultrason. Ferroelec. 
Freq.
Cont.} {\bf 39} (1992) 19.

[6] J.Y. Lu and J. F. Greenleaf, {\em IEEE Trans. Ultrason. Ferroelec. 
Freq.
Cont.} {\bf 39} (1992) 441.

[7] E. Capelas Oliveira and W. A. Rodrigues, Jr, {\em Ann. der Physik} 
{\bf 7%
} (1998) 654.

[8] P. Saari and K. Reivelt, {\em Phys. Rev. Lett}. {\bf 21} (1997) 
4135.

[9] J. E. Maiorino and W. A. Rodrigues, Jr., {\em What is Superluminal 
Wave
Motion}?, (electronic book at http://www.cptec.br/stm, {\em Sci. and 
Tech.
Mag. }{\bf 4}(2) 1999{\em ). }

[10] R. K. Sachs and H. Wu, {\em General Relativity for Mathematicians},
Spring Verlarg, New York, 1977.

[11] W.A . Rodrigues, Jr. and M. A. F. Rosa, {\em Found. Phys.} {\bf 19}
(1989) 705.

[12] W. A. Rodrigues, Jr. and E. Capelas de Oliveira, {\em Phys. Lett. 
A}
{\bf 140} (1989) 479.

[13] A. Sommerfeld, {\em Optics}, Academic Press, New York, 1952.

[14] L. Brillouin, {\em Wave Propagation and Group Velocity}, Academic
press, New York, 1960.

[15] F. A. Mehmeti, {\em Transient Tunnel Effect and Sommerfeld 
Problem},
Akademie Verlag, Berlin, 1996.

[16] G. Nimtz, {\em Ann. der Physik} {\bf 7 }(1998), 618.

[17] J. Durnin, J. J. Miceli, Jr. and J. H. Eberly, {\em Phys. Rev. 
Lett}.
{\bf 58} (1987)1499.

[18] I. S. Gradsteyn and I. M. Ryzhik, {\em Tables of Integrals, Series 
and
Products}, 4th edition, prepared by Yu.V. Geronimus and M. Yu. Tseytin,
translated by A. Jeffrey, Academic Press, New York, 1965.

[19] X. Zhou, {\em Possibility of a light pulse with speed greater than} 
$c$%
, in publ. in {\em Phys. Lett. A} (2001).

[20] R. Fox, C. G. Kuper and S. G. Lipson, {\em Proc. Roy. Soc. London 
A}
{\bf 316} (1970) 515.

[21] P. Moretti and A. Agresti, {\em N. Cimento B} {\bf 110} (1995) 905.

[22] M. E. Taylor, {\em Pseudo Differential Operators}, Princeton Univ.
Press, Princeton, 1981.

[23] R. Courant and D. Hilbert, {\em Methods of Mathematical Physics}, 
vol.
{\bf 2}, John Wiley and Sons, New York, 1966.

[24] I. N. Sneddon, {\em Fourier Transforms}, Dover Publ. Inc., New 
York,
1995

[25] J. A. Stratton, {\em Electromagnetic Theory}, McGraw-Hill, New 
York,
1941.

[26] W. K. H. Panofski and M. Phillips, {\em Classical Electricity and
Magnetism}, 2nd edition, Addison-Wesley, Reading, MA, 1962.

[27] T. T. Wu and H. Lehmann, {\em J. Appl. Phys}. {\bf 58} (1985) 2064.

\end{document}